\documentclass[convention,express-paper]{aesconf}

\graphicspath{{./}{figures/}}

\usepackage[utf8]{inputenc}

\usepackage{microtype}

\usepackage[numbers,square]{natbib}

\usepackage{booktabs}
\usepackage{color}
\usepackage{url}
\usepackage{makecell}
\usepackage{tabularx}
\usepackage{multirow}
\usepackage{array}
\usepackage{caption}
\usepackage{makecell}
\usepackage[table]{xcolor}
\usepackage{array}

\title{EMORSION – Examining the Impact of Audio Parameters on Emotional Responses and Immersion in Film.}

\author[1]{Nelly Garcia*}
\author[1]{Ruby Crocker*}
\author[1]{Bleiz M. Del Sette}
\author[1]{Fabrizio Smeraldi}
\author[1]{Charalampos Saitis}
\author[1]{George Fazekas}
\author[1]{Joshua Reiss}
\affil[*]{These authors contributed equally and share first authorship}
\affil[1]{Queen Mary University of London}

\correspondence{Nelly Garcia}{n.v.a.garcia-sihuay@qmul.ac.uk}

\lastnames{Garcia, Crocker, et al.}

\shorttitle{EMORSION - Examining the Impact of Audio Parameters in Film}

\begin{document}

\twocolumn[
\maketitle 

\begin{onecolabstract}
EMORSION is an exploratory proof-of-concept study examining how film audio design shapes audience emotion and immersion in acinema setting. Four film scenes were selected across the horror (2) and drama (2) genres, balanced between mainstream and independent productions. For each scene, multiple alternative audio mixes were created by systematically manipulating three core aspects of audio design, frequency (pitch), dynamics (loudness), and directionality (spatial placement). Three audience groups viewed the scenes, with each group exposed to one manipulated mix alongside a control mix for each scene. Audience responses were assessed through a triangulated multimodal framework combining self-reported emotion and immersion via a questionnaire, physiological measures including heart rate monitoring, and video-based motion tracking. The protocol successfully captured measurable, interpretable differences across audio conditions, indicating that even subtle changes in audio design can shape emotional perception and immersion. Unconventional mixes tended to produce greater variability in audience interpretation, while conventional immersive mixes were associated with stronger cross-audience agreement. These findings establish the feasibility of the EMORSION protocol and motivate larger-scale studies to characterise the role of specific audio parameters in shaping audience experience.


\end{onecolabstract}
]
\section{Introduction}
Sound plays a central role in shaping emotional responses and immersion in film \cite{Chion1994AudioVisionSO}, enhancing the narrative and helping communicate the director's intended message \cite{Garner}. While the role of music has been extensively studied, sound effects have received considerably less empirical attention; Kock and Louven \cite{kockPowerSoundDesign2019} showed that both music and sound design contribute significantly to perceived immersion and suspense, with their combination producing the strongest effect. Nevertheless, isolating the perceptual contribution of individual audio parameters remains methodologically challenging, particularly for sound effects, and existing work has rarely been conducted in ecologically valid contexts such as cinema environments, where the listening conditions differ substantially from typical laboratory setups \cite{greenePalgraveHandbookSound2016}. To address this gap, we introduce EMORSION (Examining the Impact of Audio Parameters on Emotional Responses and Immersion in Film), an experimental protocol designed to investigate how audio augmentation in film mixes influences audience immersion, emotional interpretation, and affective response in a cinema setting, with participants acting as part of a live audience.

In this paper, 'immersion' refers to a viewer’s state of deep mental involvement with an audiovisual experience, in which attention is strongly oriented toward the film and away from awareness of the surrounding physical environment \cite{agrawal2019defining}. Recent neuroscientific research highlights triangulation as a robust framework for studying perception, integrating physiological, behavioural, and self-report measures to capture multiple dimensions of human experience \cite{Hodges,Ronan}. A common quantification of immersion across these measures is \textit{response similarity}: the closer participants' reactions are to those observed in real-world situations, the greater the inferred level of immersion \cite{zhang2020and}. In EMORSION, immersion is not treated as a directly measurable variable; instead we adopt a triangulated perspective, drawing on subjective, physiological, and behavioural indicators to characterise audience experience, acknowledging that immersive experience is inherently personal. This study serves as a proof of concept, demonstrating that triangulated measurement is feasible in a real cinema setting and that audio modifications produce measurable effects on audience experience.

\section{Related Work}\label{related_work}

Sound design and music significantly shape audience perception, with growing research examining their role in immersive experiences \cite{Anestesis, Saroka, Crocker2024SoundStorytelling}. When film scenes are ambiguous, viewers rely heavily on music to infer mood, narrative direction, and character traits \cite{ansani2020soundtracks}. While most studies focus on individual participants in controlled laboratory settings, recent work has begun to explore cinema-like environments and real theatres in order to capture collective audience experiences \cite{Zulato, theodorou2019engaging}. 

Building on this body of work, immersion has been quantified through three complementary set of measures: subjective, physiological, and behavioural \cite{zhang2020and, jennett2008measuring}. Subjective measures rely primarily on self-report questionnaires, which remain a central method for assessing emotional expression, perception, and induction in music research \cite{Juslin}. Physiological measures capture autonomic responses linked to emotional engagement; electrocardiography (ECG), which measures the heart’s electrical activity,  is among the most widely used, and commercially available heart rate monitors such as the Polar H10 have demonstrated high accuracy and reliability in validation studies \cite{gilgen2019rr}. Furthermore, Rooney et al. \cite{rooney2014viewer} further linked decreases in heart rate to increased immersion, consistent with the idea that film immersion is an absorbed state characterised by calm focus. Behavioural measures have employed movement analysis \cite{Gonzalez}, with prior film research associating stillness and interpersonal synchrony with immersion, whereas hand movements or coughing may signal audience response or disengagement \cite{theodorou2019engaging}. Subtler cues such as nods or taps can also indicate musical immersion \cite{Ronan}, and spatialised sound can evoke orienting responses, such as head or body movements toward a sound source, that indicate attentional orientation \cite{Leman}.

Beyond the choice of measures, the temporal design of immersive experiences also matters. Immersion does not scale linearly with content duration: prior work suggests that approximately seven minutes may be optimal for spatial immersive experiences, a phenomenon referred to as duration neglect \cite{zhang2018long}. This finding directly informed the stimulus design of EMORSION, which combines subjective, physiological, and behavioural measures with bounded clip durations to support meaningful comparison across audio conditions in a live cinema setting.

\section{Methodology}

\begin{table*}[t]
\centering
\footnotesize
\caption{Selected film scenes and augmented mix assignment per session.}
\label{tab:film_scenes}
\resizebox{\textwidth}{!}{%
\begin{tabular}{|l|l|l|l|l|c|c|c|}
\hline
\rowcolor{gray!10}
\textbf{Film} & \textbf{Genre} & \textbf{Timeline} & \textbf{Duration} & \textbf{Target Emotion} & \textbf{Session 01} & \textbf{Session 02} & \textbf{Session 03}\\
\hline
Ford vs Ferrari (FVF)  & Adventure/Suspense & 2h02--2h10 & 8 min & Tense, wonder      & Dynamics      & Directionality & Frequency\\
A Quiet Place (AQP)  & Horror             & 5:00--10:00 & 5 min & Sad, tense         & Frequency     & Directionality & Dynamics\\
I Saw the TV Glow (ISTVG) & Horror            & 58:45--1h04 & 5 min & Intrigue, tense    & Frequency     & Dynamics       & Directionality\\
Decision to Leave (DTL)  & Suspense           & 1h35--1h46  & 10 min & Tense, intrigue   & Directionality & Frequency     & Dynamics\\
\hline
\end{tabular}%
}
\end{table*}

\begin{figure}
     \centering
     \includegraphics[width=0.6\linewidth]{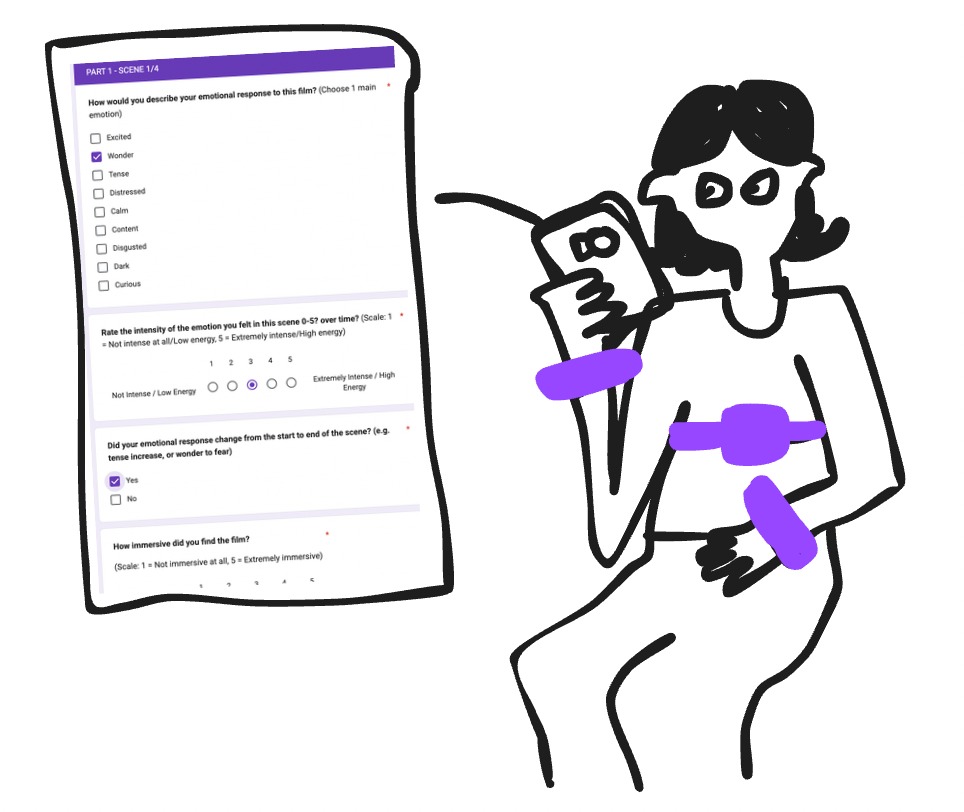}
     \caption{Participant setup illustrating behavioural tracking (reflective wristbands), physiological monitoring (sensor strap), and self-report data collection via mobile device.}
     \label{participants-fitted}
\end{figure}

We conducted three sessions at BLOC Studios, a cinema facility with a 36-speaker Dolby Atmos system and 4K projection.\footnote{https://www.qmul.ac.uk/bloc/}. In each session, participants viewed four film scenes, each presented twice — once as a control mix and once as an augmented mix (eight presentations total). Reflecting the triangulation approach, data were gathered across three modalities: physiologically, via a Polar H10 chest-strap sensor\footnote{https://www.polar.com/uk-en/sensors/h10-heart-rate-sensor} for continuous heart rate monitoring; behaviourally, via two stationary cameras capturing movement proxies such as stillness and fidgeting, supported by reflective wristbands for manual motion analysis; and subjectively, via a six-item self-report questionnaire\footnote{https://shorturl.at/EvXGO} completed on participants' mobile devices after each scene, measuring emotional response and perceived immersion. Each session opened with a 15-minute introduction covering study objectives, participant expectations, and informed consent for physiological and video data collection, and concluded with an open group discussion.

\subsection{Participants}\label{participants}

A total of 40 participants took part in the study (17 male, 22 female, and 1 non-binary). Session 1 included 13 participants (5 male, 8 female); Session 2 included 13 participants (4 male, 8 female, and one non-binary); Session 3 included 14 participants (9 male and 5 female). There was a diverse range of nationalities among participants, including English (21), European (9), Chinese (5), Mexican (2), and one participant each for Iran, India, Egypt, and Turkey.

\subsection{Film Scene Selection and Audio Modifications}

The four films scenes were selected following consultation with experts from Queen Mary School of Drama and two professional sound engineers. Selection criteria required scenes with a strong balance of music and sound effects (including Foley and environmental sound), a stand-alone narrative, and a suitable emotional range for immersive viewing. Horror and drama were chosen to limit stylistic variability while retaining within-genre diversity. Both mainstream and independent productions were included, as independent films have demonstrated comparable or greater immersive and emotional impact than mainstream counterparts \cite{aditya2024independent}. One independent and one mainstream scene were selected for each genre, and all scenes ran between 5 and 10 minutes. Prior familiarity with the selected films was low across participants, with only two or three recognising each of the films presented.

For each film scene four distinct mixes were created: an original control mix (7.1.2 Dolby Atmos) and three augmented mixes. Augmented mixes varied across three conditions: frequency, directionality, and dynamics. Each augmented mix was modified solely along its respective axis, and this resulted in a total of 16 unique audio mixes. The specific audio parameters modified for each mix are summarised and described below:

\begin{itemize}
    \item \textbf{Dynamics:} Manipulation of level and dynamic range via compressors, limiters, and expanders, controlling contrast between soft and loud events.
    \item \textbf{Frequency:} Modifications to spectral and pitch-related characteristics, brightness, timbral weight, and tonal centre, using equalization, saturation, distortion, and key transposition.
    \item \textbf{Directionality:} Alteration of spatial audio distribution via stereo and 5.1 Atmos panning, affecting sound source localisation and spatialisation.
\end{itemize}

All mixes were produced in Reaper and DaVinci Resolve using factory plug-ins. Selected scenes and timelines are presented in Table~\ref{tab:film_scenes}. Scene order and augmentation selection was counterbalanced.

\section{Results}

\begin{table*}[t]
\centering
\caption{Key self-report results by film and mix condition. Statistically significant immersion p-values in bold.}
\label{tab:SelfReport}
{%
\begin{tabular}{|c|c|c|c|c|}
\hline
\rowcolor{gray!10} \textbf{Film} & \textbf{Mix} & \textbf{Dominant Emotion} & \textbf{Immersion (p)} & \textbf{Most Salient}\\
\hline
\rowcolor{gray!20} \multicolumn{5}{|c|}{\textbf{Ford vs Ferrari (FVF)}}\\
\hline
Original & S3 & Tense (45.9\%) & \textbf{0.01} & SFX (57.1\%)\\
\hline
Frequency & S3 & Calm (28.9\%) & \textbf{0.01} & SFX (50.0\%)\\
\hline
\rowcolor{gray!20} \multicolumn{5}{|c|}{\textbf{A Quiet Place (AQP)}}\\
\hline
Original & S2 & Tense (69.2\%) & \textbf{0.002} & SFX (53.8\%)\\
\hline
Directionality & S2 & Tense (69.2\%) & \textbf{0.002} & SFX (64.3\%)\\
\hline
\rowcolor{gray!20} \multicolumn{5}{|c|}{\textbf{Decision to Leave (DTL)}}\\
\hline
Original & S3 & Tense (35.7\%) & \textbf{0.02} & SFX (50.0\%)\\
\hline
Directionality & S3 & Tense (42.9\%) & \textbf{0.02} & SFX (71.4\%)\\
\hline
\rowcolor{gray!20} \multicolumn{5}{|c|}{\textbf{I Saw the TV Glow (ISTVG)}}\\
\hline
Original & S2/S3 & Disgust (30.8--42.9\%) & \textbf{0.03 / 0.0006} & SFX / Visuals\\
\hline
Dynamics & S3 & Disgust, Distress (35.7\%) & \textbf{0.0006} & SFX (50.0\%)\\
\hline
Frequency & S2 & Disgust (38.5\%) & \textbf{0.03} & SFX (53.8\%)\\
\hline
\end{tabular}%
}
\end{table*}

Following the triangulation framework (see Section~\ref{related_work}), self-report, behavioral, and physiological data were analysed. Additional materials and secondary analysis report are available at \url{https://emorsion.netlify.app}.

\subsection{Self-Report Measures}

For each scene, participants completed a five-item questionnaire assessing emotional response and immersion, comparing original and augmented mixes. ANOVA was applied to emotional intensity ratings and salient element identification; chi-square tests assessed emotion selection and perceived emotional change over time; and a paired t-test evaluated immersion differences. Intensity change p-values were not statistically significant. Presentation order, most frequent responses, p-values, and percentages are reported in  Table \ref{tab:SelfReport}, with statistically significant values in bold.

\textbf{Ford vs Ferrari:} Session 1 audiences reported higher excitement for the dynamics mix and greater tension for the original. Dynamics were most frequently identified as the modified parameter. Sound effects were most salient in the directionality mix, while music was most salient for session 2 audiences (50\%). The frequency mix was perceived as perceptually distinct by 50\% of participants.

\textbf{A Quiet Place:} Session 1 audiences found audio effects most prominent in the original mix and music more prominent in augmented mixes. Session 2 audiences reported consistent tension across both mixes, with a statistically significant immersion change in the directionality mix. In session 3, nearly all participants (13/14) reported emotional changes with the dynamics mix, with sound effects remaining most prominent.

\textbf{Decision to Leave:} Session 1 associated the augmented mix with distress and foregrounded visuals, while the original emphasised tension, sound effects, and music, with pitch identified as the primary modification. Session 2 found the original mix evoked calmer, darker emotions, whereas the dynamics mix increased curiosity; 46.2\% identified pitch as the main modification. In session 3, sound effects were dominant (71.4\%) and the directionality mix showed a statistically significant immersion change ($p = 0.02$).

\textbf{I Saw the TV Glow:} Emotional patterns were broadly similar across mixes. Session 1 foregrounded sound effects; session 2 reported disgust most frequently in the frequency mix (38.5\%); session 3 found the dynamics mix eliciting disgust and distress (35.7\%), with the original predominantly eliciting disgust, and visuals and audio effects most noticeable.

Notably, the self-reported questionnaire revealed a significant effect on perceived immersion. Across all sessions, most participants reported increased immersion for at least one augmented mix compared to the original. Specifically, frequency-based mixes (ISTVG and FVF) and directionality mixes (DTL and AQP) were associated with higher immersion ratings, as reflected in the corresponding immersion-change p-values.

\subsection{Physiological Data}

The Polar H10 sensor recorded heart rate (HR, bpm) and RR intervals (ms) at 1 Hz for each participant during each scene. One participant was excluded for declining to wear the sensor. Each remaining participant viewed 8 scenes (4 control, 4 augmented). We first excluded incomplete recordings due to mid-session dropouts and removed physiologically implausible values (HR: 46--200 bpm; RR: 300--1300 ms) \cite{bruinDetectionArousalValence2024}. Afterwards, we performed interpolation via Piecewise Cubic Hermite Interpolating Polynomial \cite{benchekrounImpactMissingData2023} on all time series and further excluded windows where interpolation exceeded 30\% of samples.

Time- and frequency-domain metrics were then calculated for each remaining series (see Table~\ref{Polar H10 metrics}) \cite{bahameishStrategiesReliableStress2024a, bruinDetectionArousalValence2024}. Data were aggregated into 12 subsets by pairing each augmented condition with its corresponding control, and pairwise t-tests were conducted per metric with Benjamini-Hochberg correction for multiple comparisons.

Frequency-domain measures yielded no significant results after correction, except a marginal effect for total power in the DTL directionality mix ($p = 0.035$). Time-domain metrics were more informative: SDNN showed significant effects for dynamic mixes of DTL ($p = 0.014$) and AQP ($p < 0.001$); HR standard deviation similarly for DTL ($p = 0.035$) and AQP ($p < 0.001$), with an additional effect for the DTL frequency mix ($p = 0.035$). HR interquartile range showed significant effects for dynamic mixes of DTL ($p = 0.04$), AQP ($p = 0.01$), and ISTVG ($p = 0.01$).

\begin{table*}[t]
\centering
\footnotesize
\begin{tabularx}{0.91\textwidth}{c|l}
\hline
\rowcolor{gray!20} \multicolumn{2}{c}{\textbf{RR Time Domain Metrics}}\\
\textbf{SDNN} & Standard deviation of RR intervals \\
\textbf{RMSSD} & Root mean square of successive differences, \\
\textbf{RR Mean} & Mean of RR intervals \\
\textbf{pNN20 and pNN50} & Proportion of successive pairs that differ by more than 20/50 ms divided by total number of RR \\
\hline
\rowcolor{gray!20} \multicolumn{2}{c}{\textbf{HR Time Domain Metrics}}\\
\textbf{Mean and Median HR} & Overall mean and median values for HR \\
\textbf{HR STD} & Standard deviation of HR \\
\textbf{IQR HR} & Interquartile Range (i.e., range of the middle 50\% of measurements) of HR \\
\textbf{Mean Difference} & Mean difference of HR \\
\rowcolor{gray!20} \multicolumn{2}{c}{\textbf{RR Frequency Domain Metrics}}\\
\textbf{VLF, LF, and HF Power} & Power in the very low (0-0.04 Hz), low (0.04-0.15 Hz), and high (0.15-0.4 Hz) frequencies \\
\textbf{Total Power} & Sum of the power in all frequency bands \\
\textbf{LF/HF} & Ratio between low frequencies and high frequencies \\
\hline
\end{tabularx}
\caption{Description of the extracted metrics from RR and HR measurements}
\label{Polar H10 metrics}
\end{table*}

\subsection{Movement Tracking} \label{movement-results}

\begin{figure}
    \centering
    \includegraphics[width=1.0\linewidth]{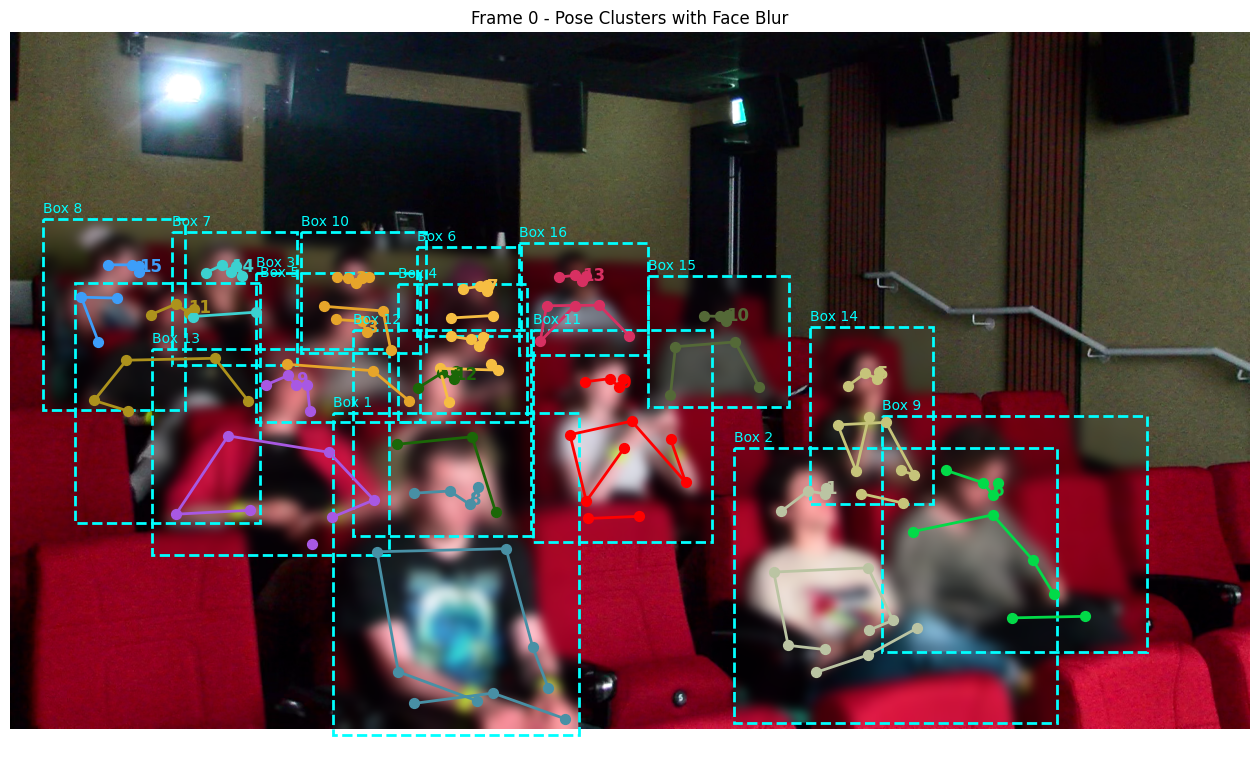}
    \caption{Pose skeletal keypoints for Movement detection of participants with bounding boxes}
    \label{pose-cluster}
\end{figure}

Audience movement was analysed using video-based motion tracking (Figure~\ref{pose-cluster}). Skeletal movement was extracted using OpenPose via OpenPifPaf \footnote{https://openpifpaf.github.io/intro.html}, which estimates 2D body keypoints with associated confidence scores. Only keypoints exceeding a fixed confidence threshold (above \%10) were included and assigned to participants using manually defined bounding boxes corresponding to seating positions. To reduce noise and computational load, analysis was performed on temporally subsampled frames at approximately 1 Hz ($~480$ frames for an 8-minute scene). Out of focus, blurred or excessively low light footage was excluded from the analysis. Total movement was quantified as the sum of frame-to-frame skeletal keypoint displacements across the scene, normalised by bounding-box size, weighted by keypoint confidence, and aggregated to produce a per-participant movement magnitude. Additionally, we computed four metrics to capture other salient aspects of the viewing experience. \textbf{Mean and SD Movement} summarise total movement across participants - in other words, how active the audience was during the scene: a high mean indicates that participants moved more overall, while a low mean indicates that they remained largely still, and the standard deviation captures between-participant variability. \textbf{Mean and SD Synchrony} track the pairwise alignment of participants' movements, measured via cosine similarity of skeletal vectors on a scale from 0 to 1, where 1 indicates strong synchrony and 0 indicates none; the standard deviation reflects variability in alignment across participants. The most salient results are shown in Table~\ref{tab:movement_summary}, a complete summary of the analysis can be found at \url{https://emorsion.netlify.app}.

Across all scenes, mean audience movement ranged from 122 to 535 (global mean = 330). Horror scenes (ISTVG, AQP) consistently elicited lower-than-average mean movement, indicating group physical stillness across participants, whereas FVF and DTL produced higher movement levels overall. Temporal variability in movement (SD Movement) differed across films, reflecting whether activity was sustained or concentrated in brief moments. Overall movement levels differed across films and sessions, though this pattern was not consistent across all stimuli. Synchrony values were generally higher during second viewings, suggesting increased collective alignment over time. Directionality mixes increased mean movement in comparison to the original mixes.

Despite relatively high overall movement, \textit{FVF} showed stable synchrony and low variation between participants (0.81-0.83), suggesting that audiences tended to move in similar ways. By contrast, \textit{ISTVG} exhibited lower movement (0.000-0.110) with higher synchrony (0.86), consistent with a more still and collectively absorbed viewing style, reinforcing participants claiming to be more immersed. Between-participant variability (SD Movement) was also high (SD = 349.03), suggesting that some audience members moved extensively while others remained comparatively still.

\begin{table*}[t]
\centering
\footnotesize
\caption{Summary of head movement and synchrony by film and mix.}
\label{tab:movement_summary}
\begin{tabular}{|c|c|c|c|c|c|} 
\hline
\rowcolor{gray!10} \textbf{Mix} & \textbf{Mean Move.} & \textbf{SD Move.} & \textbf{Rel. Activity} & \textbf{Mean Sync.} & \textbf{SD Sync.} \\
\hline
\rowcolor{gray!20} \multicolumn{6}{|c|}{\textbf{Ford vs Ferrari (FVF)}}\\ 
\hline
Original & 1.000 & 349.03 & Very High & 0.83 & 0.02 \\
\hline
Dyn & 0.814 & 353.65 & High & 0.81 & 0.02 \\
\hline
\rowcolor{gray!20} \multicolumn{6}{|c|}{\textbf{A Quiet Place (AQP)}}\\ 
\hline
Original & 0.396 & 245.23 & Below Avg. & 0.81 & 0.04 \\
\hline
Dir & 0.404 & 240.33 & Below Avg. & 0.77 & 0.05 \\
\hline
\rowcolor{gray!20} \multicolumn{6}{|c|}{\textbf{Decision to Leave (DTL)}}\\ 
\hline
Original & 0.928 & 487.92 & High & 0.81 & 0.03 \\
\hline
Dyn & 0.905 & 512.79 & High & 0.81 & 0.04 \\
\hline
\rowcolor{gray!20} \multicolumn{6}{|c|}{\textbf{I Saw the TV Glow (ISTVG)}}\\ 
\hline
Original & 0.019 & 111.31 & Very Low & 0.82 & 0.07 \\
\hline
Freq & 0.110 & 165.30 & Low & 0.82 & 0.03 \\
\hline
\end{tabular}
\end{table*}

\section{Discussion}

In this study we combined self-report, physiological, and behavioural measures, using a triangulated method to assess immersion in a live cinema setting. 

Self-report provided the strongest evidence: participants reported higher immersion for at least one augmented mix per scene (see Table~\ref{tab:SelfReport}). Directionality manipulations had the greatest impact for DTL and AQP. For AQP, increased spatialisation aligned with reduced movement; for DTL, a slight heart rate increase suggested heightened engagement, with sound effects most frequently reported as salient. This supports prior work showing surround sound can function as a focusing mechanism, drawing attention inward through enhanced spatial cues \cite{mendoncaSurroundSoundSpreads2020}.

Physiological measures were most sensitive to dynamics mixes across DTL, AQP, and ISTVG. AQP showed reduced heart rate variability alongside self-reports of stress and tension, suggesting sustained arousal. DTL and ISTVG exhibited heart rate increases, indicating heightened reactivity rather than a uniform response, consistent with evidence that dynamically intense audio can elicit aversive responses \cite{dimitrievEffectAuditoryStimulation2023}. For ISTVG, increased movement co-occurred with higher immersion ratings and stronger reports of \textit{nervousness} and \textit{disgust}, suggesting movement reflected emotional discomfort rather than disengagement \cite{ottavianiAutonomicCorrelatesPhysical2013, kreibigAutonomicNervousSystem2010, kurakata2013sensory, ringer2024sounds}.

DTL, AQP, and ISTVG are all tension- or suspense-driven scenes, and all three showed measurable changes across mix conditions. This suggests suspense-driven films are particularly well suited to multimodal immersion studies, though dominant scene emotions (e.g.\ disgust in ISTVG) should be carefully considered in both design and analysis.

The FVF frequency mix increased self-reported immersion without corresponding physiological or movement changes, suggesting limited emotional impact. Perceptual biases such as peak and recency effects may have shaped ratings, as gradual frequency changes are less likely to register as salient moments \cite{zhang2018long}.

Overall, subtle audio augmentations, particularly dynamics and directionality, can meaningfully influence subjective engagement, though different parameters elicit responses in different ways depending on the film.

\section{Limitations and Future Work}

Despite these initial encouraging results, limitations should be acknowledged in the current data collection pipeline and stimulus design. Movement tracking proved the most challenging measure: one session yielded almost no usable data, restricting cross-session comparisons, and frame rate reduction to 1 fps limited temporal resolution. Stimuli selection and mix precision were constrained by limited access to commercial audio stems, while sensor availability and camera coverage restricted sample size and tracking accuracy at peripheral seating positions.

Future work should address these constraints along two complementary directions. First, integrating qualitative movement data could strengthen the empirical links between movement analysis and experiential states, while a multi-camera or infrared sensor array would improve tracking reliability in the low-light conditions typical of cinema. Second, in terms of stimuli, securing stem access through student or independent productions would enable more precise manipulation, and parameters beyond volume and spatialisation - such as timbre via synthesis or machine-learning transfer - could be explored; collaboration with professional mixers would also help ensure that augmented mixes meet industry standards.

\section{Conclusions}

The findings of our proof of concept confirm the validity of the triangulation method, from which we observed patterns and results consistent with those reported in previous controlled studies. Furthermore, we demonstrate that subtle changes in the audio domain can be effectively explored with this approach to assess their influence in an ecologically valid environment. The multimodal analysis showed that self-reported immersion was most sensitive to audio manipulations, while physiological and behavioural measures provided complementary evidence whose variability itself offers insight into the heterogeneity of audience response. Together, these results indicate that the EMORSION protocol is sufficiently sensitive to detect the effects of subtle audio manipulations on audience experience, motivating larger-scale studies to characterise how specific audio parameters shape narrative engagement.


\footnotesize 
\bibliographystyle{jaes}

\bibliography{refs}

\end{document}